\documentclass[useAMS,usenatbib]{mn2e}

\usepackage{amssymb}
\usepackage{amsfonts}
\usepackage{mncite}
\usepackage{epsfig}
\usepackage{psfig}

\begin{document}

\title[Self-gravitating disc structure]
{Time-dependent models of the structure and stability of self-gravitating protoplanetary discs.}

\author[W.K.M. Rice \& Philip J. Armitage]
 {W.K.M. Rice$^1$\thanks{E-mail: wkmr@roe.ac.uk}, Philip
 J. Armitage$^{2,3}$ \\
$^1$ SUPA\thanks{Scottish Universities Physics Alliance},
Institute for Astronomy, University of Edinburgh, Blackford Hill, Edinburgh, EH9 3HJ \\
$^2$ JILA, Campus Box 440, University of Colorado, Boulder CO 80309 \\
$^3$ Department of Astrophysical and Planetary Sciences, University of
 Colorado, Boulder CO 80309 \\}

\maketitle

\begin{abstract}
Angular momentum transport within young massive protoplanetary discs may 
be dominated by self-gravity at radii where the disk is too weakly 
ionized to allow the development of the magneto-rotational instability. 
We use time-dependent one-dimensional disc models, based on a local 
cooling time calculation of the efficiency of transport, to study the radial structure 
and stability (against fragmentation) of protoplanetary discs in which 
self-gravity is the sole transport mechanism. We find that self-gravitating 
discs rapidly attain a quasi-steady state in which the surface density 
in the inner disc is high and the strength of turbulence very low 
($\alpha \sim 10^{-3}$ or less inside 5~au). Temperatures high enough 
to form crystalline silicates may extend out to several au at early 
times within these discs. None of our discs spontaneously develop 
regions that would be unambiguously unstable to fragmentation into substellar 
objects, though the outer regions (beyond 20~au) of the most massive discs 
are close enough to the threshold that fragmentation cannot be ruled out. 
We discuss how the mass accretion rates through such discs may 
vary with disc mass and with mass of the central star, and note 
that a determination of the $\dot{M}$-$M_*$ relation for very young 
systems may allow a test of the model.
\end{abstract}

\begin{keywords}
stars: formation --- stars: pre-main-sequence --- circumstellar matter --- planetary systems: protoplanetary discs ---
planetary systems: formation
\end{keywords}

\section{Introduction}
Low mass stars form from the collapse of cold, dense molecular cloud cores \citep{terebey84}. 
Although the rotation rates of such cores are generally quite small \citep{caselli02} they nonetheless contain 
amounts of angular momentum far in excess of the rotational angular momentum of a single 
star. Most of the mass must therefore pass through a protostellar disc, and the answers to 
many open problems in star and planet formation hinge on the nature of the angular 
momentum transport that is needed for disc accretion.

In most astrophysical discs the fundamental question of what mechanism dominates 
angular momentum transport is widely considered to have been solved -- MHD turbulence 
initiated by the magneto-rotational instability (MRI) can provide the
necessary viscosity \citep{balbus91,papaloizou03}. Protoplanetary disks, however, 
are so cold and dense that thermal processes probably fail to yield even the 
very small degree of ionization needed to sustain MHD turbulence\citep{blaes94}. 
Under these conditions disc self-gravity may provide an alternate and possibly dominant 
mechanism for transporting angular momentum through the growth of the gravitational 
instability \citep{toomre64,lin87,laughlin94}.

Study of the development of gravitational instability in protostellar discs has 
often been motivated largely by the possibility that the instability will lead to 
fragmentation of the disc and the formation of gas giant planets \citep{boss98}.
The conditions for fragmentation are, however, quite difficult to achieve, especially in the
inner, planet-forming regions \citep{matzner05,rafikov05,boley06,whitworth06,stamatellos08,forgan09}. What seems more 
likely is that discs will evolve towards quasi-steady states in which the instability 
acts to transport angular momentum outwards \citep{gammie01,rice03,lodato04,vorobyov07}. This is 
of interest in its own right, as it implies that the conditions within young 
protoplanetary discs -- at precisely the epoch when planetesimals and perhaps larger 
bodies are forming -- may largely be set by the physics of angular momentum 
transport via gravitational instability. The recent progress in understanding the 
conditions for fragmentation has yielded a much clearer understanding of this 
physics, which we use here to construct realistic time-dependent models 
for the structure of self-gravitating protoplanetary discs. Our models rely 
on two specific properties of transport via gravitational instability. First, 
that transport can be approximated as a local viscous process for all except 
the most massive discs \citep{lodato04,lodato05}. Second, under conditions of thermal equilibrium 
the strength of angular momentum transport is set by the cooling rate 
of the disc \citep{gammie01,rice03}. Our results show that self-gravitating discs will settle 
into quasi-steady states that appear independent of the initial conditions, 
but that their properties are quite unlike those that result from angular 
momentum transport via generic turbulent processes. Specifically, we 
find that the surface density profile
is reasonably steep and in the cases considered here, $\sim 80$ \% of the mass within $50$ au 
is located inside $10 - 20$ au. The quasi-steady mass accretion rate depends strongly on
the disc mass and on the mass of the central star.  For a constant
star to disc mass ratio, however, the relationship
between mass accretion rate and central star mass is similar to that
found observationally.We further show that the secular evolution of the 
disc does not typically result in {\em any} regions that would be unstable to 
fragmentation, with the region inside $10 -20$~au being particularly stable unless some mechanism - such 
as convection \citep{boss04} - can significantly increase the cooling rate. 

Our focus in this paper is on the outer cool regions of protoplanetary discs, 
and for this reason and for simplicity we consider disk models in which 
self-gravity provides the sole source of angular momentum transport. 
Of course both the extreme inner region (inside 0.1~au) and the upper 
layers of the disc further out could become sufficiently ionised for the MRI 
to operate \citep{gammie96}, and this would modify our results and admit 
new physical effects. In particular, a pile-up of material brought to within $1-2$~au 
by self-gravity could eventually trigger the onset of the MRI and 
episodic outbursts \citep{armitage01,zhu09}. These might be related to 
the FU Orionis phenomenon \citep{hartmann96}.    Ultimately the evolution of discs 
around very young stars could be driven by both
MRI and the gravitational instability \citep{terquem08}.

The paper is organised as follows. In Section 2 we describe how we can
self-consistently model protostellar discs evolving through
self-gravity alone.  In Section 3 we describe our results, and in
Section 4 we summarise our conclusions. 

\section{Disc models}
We model the evolution of a self-gravitating protoplanetary disc under 
the assumptions that the potential is fixed, that the disc is in thermal 
equilibrium at all radii, and that the local strength of angular 
momentum transport is set by self-gravity.

\subsection{Viscous evolution}
The surface density $\Sigma(r,t)$ of an axisymmetric disc 
evolves according to \citep{lynden74,pringle81}
\begin{equation}
\frac{\partial \Sigma}{\partial t} = \frac{3}{r} \frac{\partial}{\partial r}\left[ r^{1/2}
\frac{\partial}{\partial r} \left( \nu \Sigma r^{1/2} \right) \right],
\label{1Ddiff}
\end{equation}
where $\nu$ is the kinematic viscosity. If the disc is able to 
attain a steady-state this equation can be integrated to give an 
expression for the mass accretion rate $\dot{M}$ which, at radii large 
compared to the radius of the star, is 
\citep{pringle81}
\begin{equation}
\dot{M} = 3 \pi \nu \Sigma.
\label{Mdot}
\end{equation} 
The steady-state radial surface density 
profile is therefore determined by the radial profile of the viscosity. 
We adopt the $\alpha$ formalism
\citep{shakura73} in which the viscosity is taken to be 
$\nu = \alpha c_s H$ where $c_s$ is the
disc sound speed, $H = c_s/\Omega$ is the disc scale height, and 
$\alpha << 1$ is a 
parameter that determines the efficiency of angular momentum transport.

The strength of self-gravity depends upon the thermal properties of the 
disc. In thermal equilibrium the viscosity generates dissipation in the 
disc at a rate $D(R)$
per unit area per unit time, where \citep{bell94}
\begin{equation}
D(R) = \frac{9}{4} \nu \Sigma \Omega^2,
\label{diss}
\end{equation}
which must balance cooling via thermal emission from the disc 
surfaces (if the disc is optically thick) or via optically thin 
emission integrated vertically through the disc. 

\subsection{Determination of $\alpha$}
An accretion disc can become gravitationally unstable if \citep{toomre64}
\begin{equation}
Q = \frac{c_s \kappa}{\pi G \Sigma} \sim 1,
\label{Q}
\end{equation}
where $\kappa$ is the epicyclic frequency which is replaced by the
angular frequency, $\Omega$, in a Keplerian disc.  One possible outcome
is that unstable discs fragment to produce bound objects and has been suggested
as a possible mechanism for forming giant planets \citep{boss98,boss02}. 
For axisymmetric instabilities this requires $Q < 1$, while for 
non-axisymmetric instabilities this can occur for $Q$ values as high 
as $1.5 - 1.7$ \citep{durisen07}.  It has, however, been realised recently that the
above condition is not sufficient to guarantee fragmentation.  The disc
must also be able to lose the energy generated by the
instability \citep{gammie01,rice03}.  The rate at which energy
must be lost depends on the equation of state \citep{rice05}, but 
in protostellar discs, the cooling
time would generally need to satisfy $t_{\rm cool} \le 3 \Omega^{-1}$
\citep{gammie01}.

For cooling times greater than that required for fragmentation, the disc
will settle into a quasi-steady state in which the instability acts to
transport angular momentum \citep{laughlin96,gammie01,lodato04}. 
In principle this transport need not be local \citep{balbus99}, 
and if this were the case then neither writing  
$\alpha$ as a function of local conditions nor using equation~\ref{1Ddiff} 
would be valid. Simulations, however, show that the local approximation 
is surprisingly good at capturing the behavior of self-gravitating protoplanetary 
discs, which {\em can} therefore be regarded as having an effective viscosity 
of the form $\nu = \alpha c_s H$. Moreover, under conditions of thermal 
equilibrium the value of $\alpha$ can be derived via a simple energy 
balance requirement. 

The disc cools at a rate \citep{pringle81,johnson03}
\begin{equation}
\Lambda = 2 \sigma T_e^4,
\label{cool1}
\end{equation}
where $\sigma$ is the Stefan-Boltzmann constant, $T_e$ is the effective
temperature, and the factor of $2$ comes from the radiation escaping on
both sides of the disc.  If the disc is optically thick, radiation transport
can be treated in the diffusion approximation, and the effective temperature
can be shown to be given by \citep{hubeny90}
\begin{equation}
T_e^4 = \frac{8}{3}\frac{T_c^4}{\tau},
\label{hubeny}
\end{equation}
where $T_c$ is the temperature of the disc midplane, and $\tau$ is the 
Rosseland mean optical depth.  The internal energy per unit area, $U$, is then
\begin{equation}
U = \frac{c_s^2 \Sigma}{\gamma ( \gamma - 1)},
\label{inten}
\end{equation}
where $\gamma$ is the specific heat ratio. 

The cooling time is then  simply $t_{\rm{cool}} = U / \Lambda$, where $\Lambda$ is the cooling
function given by equation (\ref{cool1}). Since we are assuming that the disc is
in a quasi-steady state, the cooling must be balanced by dissipation.  If we
assume that the energy is dissipated locally, which appears to be
the case for $Q \sim 1$ \citep{balbus99,lodato04}, we can use equation (\ref{diss}) to
get \citep{gammie01}
\begin{equation}
\alpha = \frac{4}{9 \gamma (\gamma - 1) t_{\rm{cool}} \Omega}.
\label{alpha}
\end{equation}
We should acknowledge that the nature of the transport - whether local or global -
may depend on the form of the cooling \citep{mejia05,durisen07} and that our use
of the local approximation is clearly a simplication.  There may well be situations
in which the energy is not dissipated locally and although the disc as a whole may be
in thermal equilibrium, equation (\ref{alpha}) is not strictly valid.

\subsection{Modelling a fully self-gravitating disc}
To consider the evolution of a disc in which angular momentum transport is governed
by self-gravity alone, we assume that the disc settles into a quasi-steady state and
that energy is dissipated locally.  We assume that in this quasi-steady state the 
Toomre $Q$ parameter satisfies $Q = 1.5$, unless $T_c < 10$ K in which case we set 
$T_c = 10 K$.  For a given surface density, equation (\ref{Q}) can be used to determine
the midplane sound speed, $c_s$.  The midplane temperature can then be determined using
\begin{equation}
T_c = \frac{\mu m_p c_s^2}{\gamma k_B},
\label{midTemp}
\end{equation}
where $\mu = 2.4$ is the mean molecular weight, $m_p$ is the proton mass, and $k_B$ is
Boltzmann's constant.

\begin{figure}
\psfig{figure=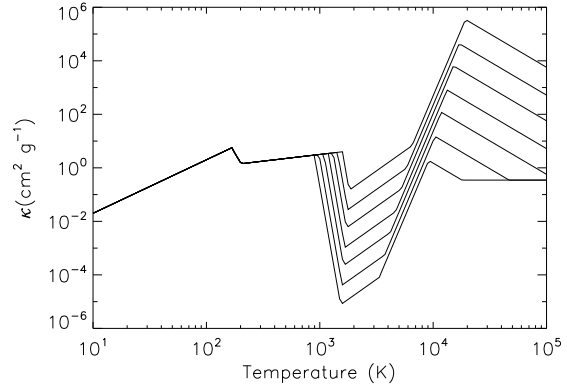, width = 0.45\textwidth}
\caption{Opacities from \citet{bell94} against temperature for densities of 
$\rho = 10^{-9}, 10^{-8}, 10^{-7}, 10^{-6}, 10^{-5}, 10^{-4}, 10^{-3}$ g cm$^{-3}$.}
\label{kappa}
\end{figure}

For a given surface density, $\Sigma$, and using the assumption that $Q = 1.5$ (unless $T < 10$K) 
we can now determine the midplane temperature.  To determine the cooling rate, however,
we need to determine the effective temperature, $T_e$, for which we need the optical depth. 
We can approximate the optical depth using
\begin{equation}
\tau = \int_0^\infty dz \kappa ( \rho_z,T_z ) \rho_z \approx H \kappa ( {\bar{\rho}}, \bar{T} ) 
\bar{\rho},
\label{tau}
\end{equation}
where $\kappa$ is the Rosseland mean opacity and $\bar{\rho}$ and $\bar{T}$ are an 
average density and temperature for which we use $\bar{\rho} = \Sigma/(2H)$ and
$\bar{T} = T_c$.  For the Rosseland mean opacities we use the analytic approximations
from \citet{bell94}.  Figure \ref{kappa} shows the opacity against temperature for a range
of densities. The opacity is dominated, in order of increasing temperature, by ice grains,
metal grains, Molecules, H$^-$ scattering, Bound-free and free-free absorption, and electron
scattering.  The sudden changes in opacity at $T \approx 170$ K and $T \approx 1000$ K
are due to the evaporation of the ice mantles on the ices grains and the evaporation of the
metal grains respectively.  

Once the optical depth, $\tau$, is known the cooling function, $\Lambda$, 
is calculated using a modified form of equations (\ref{hubeny}) and (\ref{cool1}),
\begin{equation}
\Lambda = \frac{16 \sigma}{3} (T_c^4 - T_o^4)\frac{\tau}{1 + \tau^2}
\label{modCoolFunc}
\end{equation}
where $T_o = 10$K is assumed to come from some background source that prevents the
midplane cooling below this value \citep{stamatellos07b}, and the last term is introduced to smoothly 
interpolate between optically thick and optically thin regions \citep{johnson03}.
The cooling time is then $t_{\rm cool} = U / \Lambda$ with $U$ given
by equation (\ref{inten}) and the viscous $\alpha$ can then be 
determined using equation (\ref{alpha}). The viscosity is then
$\nu = \alpha c_s H$ and the disc is evolved using equation (\ref{1Ddiff}).

\section{Results}
\subsection{Surface density profiles}
The simulations presented here assume, initially, a rather extreme case of a 
central star with a mass
of $0.35$ M$_\odot$ surround by a circumstellar disc with a mass of
$0.35$ M$_\odot$.  The disc is assumed to have an initial 
power-law surface density profile
$\Sigma \propto r^{-\beta}$ between 0.5 and 50 au.  
The assumption that $Q = 1.5$ then determines
the disc midplane temperature, $T_c$, with the additional constraint that $T_c \ge 10$ K.
The process described above determines $\alpha$ and the disc is then
evolved in time using equation (\ref{1Ddiff}).  We consider 3 different
initial surface density profile, $\beta = 2.0$, $\beta = 1.5$, and $\beta = 1.0$.  

\begin{figure}
\begin{center}
\psfig{figure=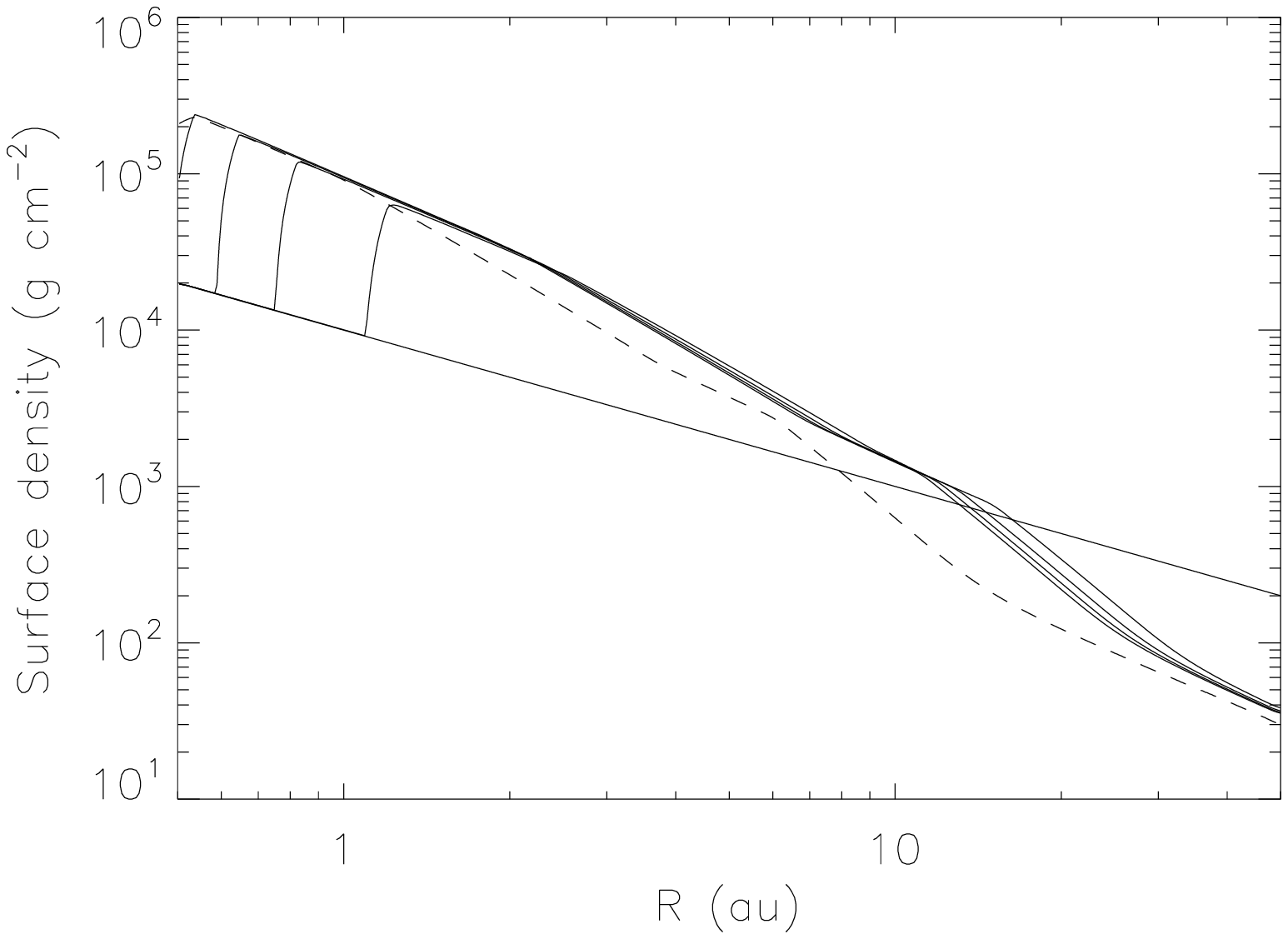,width=0.45\textwidth}
\psfig{figure=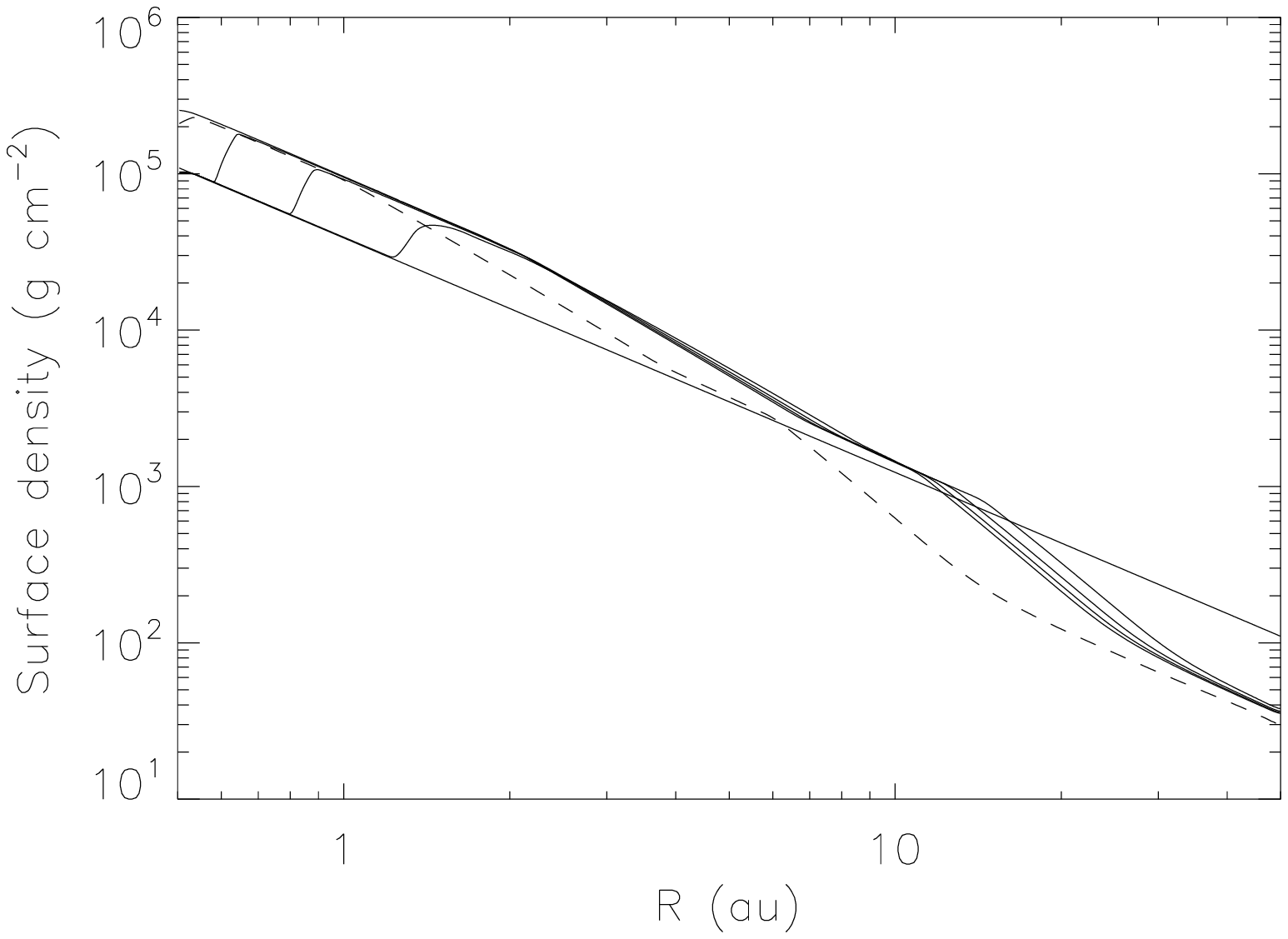,width=0.45\textwidth}
\psfig{figure=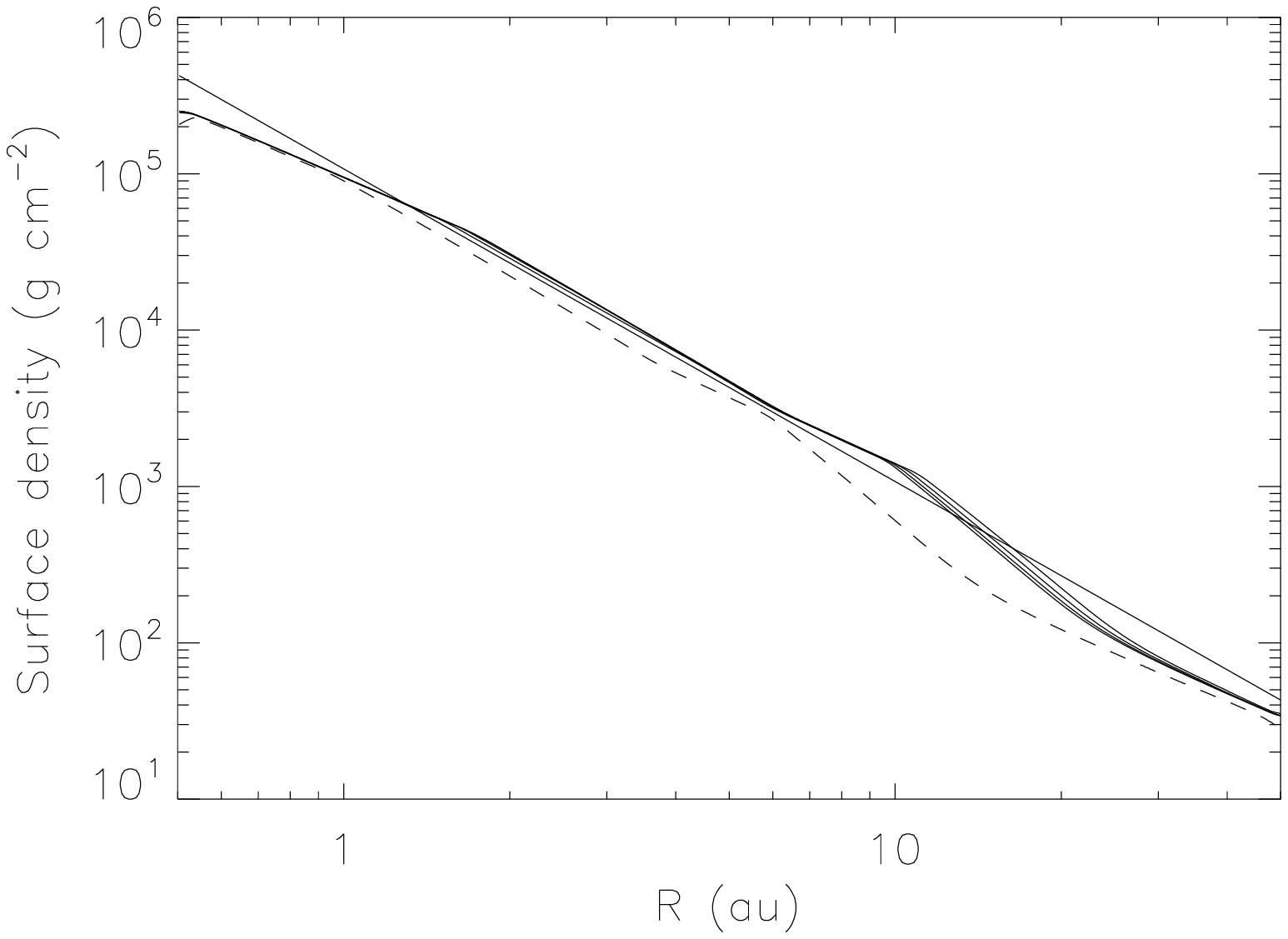,width=0.45\textwidth}
\caption{Figure showing how $0.35$ M$_\odot$ discs around a $0.35$ M$_\odot$ star with 
initial surface density profiles of $\Sigma \propto r^{-1}$ 
(top panel), $\Sigma \propto r^{-1.5}$ (middle panel), and $\Sigma \propto r^{-2}$ (bottom panel) evolve into
quasi-steady states.  In all the cases the resulting quasi-steady profiles are the
same.}
\label{sdens}
\end{center}
\end{figure}

\begin{figure}
\begin{center}
\psfig{figure=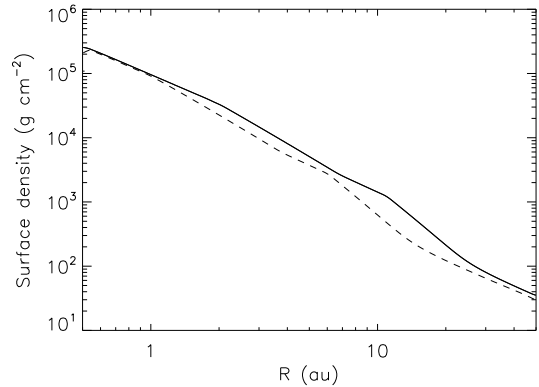,width=0.45\textwidth}
\caption{Figure showing the quasi-steady surface density profile for
a $0.35$ M$_\odot$ disc around a $0.35$ M$_\odot$ star (solid line). The 
profile moves inwards with time (dashed line) as mass accretes onto the
central star.}
\label{single_profile}
\end{center}
\end{figure}

The surface density evolution for the three 3 initial profiles is shown in Figure \ref{sdens}. We
should stress that we do not assume that the initial profiles are necessarily realistic, but simply
want to if establish if such discs will settle into quasi-steady states and if such a state depends in any way
on the initial surface density profiles.
Each panel in Figure \ref{sdens} shows the initial power-law surface density, $4$ subsequent surface density
profiles each separated by $20000$ years and the final surface density (dashed-line) at the
end of the simulation which we stop after $10^6$ years. 
What this shows is that
the surface density adjusts in all three cases to a state that appears to be largely 
independent of the initial surface density profile.  For $\beta = 1.0$ and $\beta = 1.5$
this settling takes a some time ($\sim 80000$ years), while for $\beta = 2$, it occurs
almost immediately.  For clarity, Figure \ref{single_profile} shows the surface density
profile when a quasi-steady state has just been achieved (solid line).  
With time, this profile moves inwards and the disc mass decreases, as illustrated by the dashed line in 
Figure \ref{single_profile} which shows the surface density profile when  $M_{\rm disc} = 0.25$ M$_\odot$.

\begin{figure}
\begin{center}
\psfig{figure=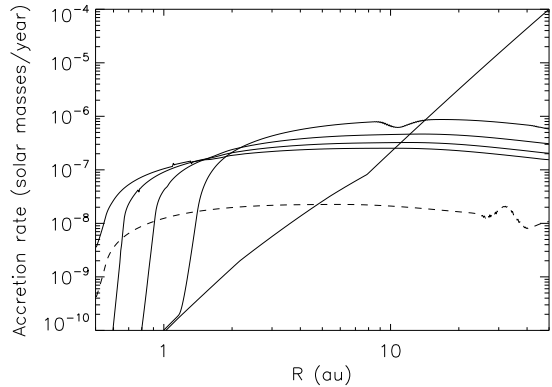,width=0.45\textwidth}
\caption{Figure showing the mass accretion rate against radius at different times for $\Sigma \propto r^{-1.5}$.
The almost diagonal line shows the initial mass accretion rate while the other solid lines show the accretion rate
at $20000$ year intervals until a quasi-steady rate is achieved with a mass accretion rate of $\dot{M} \sim 10^{-7}$
M$_\odot$/yr.  The dashed line shows the accretion rate after $10^6$ years when the disc mass 
has decreased to $0.25$ M$_\odot$. The two other initial surface density profiles (which we don't show here) evolve to the same
quasi-steady accretion rate, but over slightly different timescales, with the $\Sigma \propto r^{-2}$ disc evolving
to a quasi-steady state much more rapidly than $\Sigma \propto r^{-1}$ and $\Sigma \propto r^{-1.5}$ discs.}
\label{Mdot_fig}
\end{center}
\end{figure}

In all cases what is happening is that mass is redistributing itself
to produce a state in which the accretion rate, $\dot{M}$, is largely independent of $r$.  
The is illustrated in Figure \ref{Mdot_fig} which shows the evolution of the mass accretion
rate for $\beta = 1.5$. The behaviour is essentially the same for
$\beta = 1$ and $\beta = 2$ except that the latter reaches an 
approximately constant mass accretion rate more rapidly than the other two cases. 
In the prescription presented here, the viscous $\alpha$ depends on the local cooling rate 
- which depends on the local temperature -
and the accretion rate depends on the local viscosity and surface density (e.g., equation (\ref{Mdot})).  
For initially flatter surface density profiles ($\Sigma \propto r^{-1}$ and $\Sigma \propto r^{-1.5}$) the outer disc 
contains most of the mass. To be gravitationally stable (see equation \ref{Q}), 
the temperature in the outer disc also 
needs to be high.  The inner disc, on the other hand, can have relatively low temperatures and 
remain gravitationally stable. The cooling time, and consequently viscosity, in the outer disc is 
therefore significantly greater than in the inner disc, producing an initially much larger mass
accretion rate in the outer disc than in the inner disc. This process, however, moves mass
inwards and in doing so not only increases the surface density in the inner disc, but also the temperature to keep
the disc gravitationally stable. This increases the cooling rate in the inner disc, the viscosity, and the
mass accretion rate.  Eventually the mass is redistributed such that the entire
disc has the same mass accretion rate. If the initial surface density profile is, however, initially
close to the quasi-steady profile - as in the third panel of Figure \ref{sdens} - very little 
redistribution is required and the disc very quickly settles into a quasi-steady state. 

This mass redistribution leads to quasi-steady surface density profiles
that are reasonably steep ($\Sigma \sim r^{-2}$) and that have
quite a substantial break at 10 - 20 au.  Even though the discs extend
to $50$ au, most of the mass ($> 80$ \%) is actually located in the
inner disc. It is now generally accepted that discs disappear
within a few million years \citep{haisch01} and consequently that gas
giant planets need to form within this timescale. A substantial
reservoir
of mass in the inner disc could significantly accelerate planet growth
if gaseous planets form via the standard core accretion scenario
\citep{pollack96}. This type of profile is also qualitatively
consistent
with observations of massive discs \citep{rodriguez05,carrasco-gonzalez09} that suggest
that such discs have small radii ($r_{\rm disc} \sim 25$ au).  It is also quite likely
that with so much mass located in the optically thick inner regions, and the
sub-mm flux largely determined by the cold, outer disc, 
the masses of such discs could be easily underestimated \citep{hartmann06}.  
Such gas rich inner disc could be detected using 
near-IR gas emission lines such as the CO bandhead \citep{thi05,najita07}. However,
for the disc masses considered here, these lines
are still optically thick and will still not provide accurate probes
of the disc mass.

Although the goal here is to illustrate the quasi-steady structure of a self-gravitating disc,
it is also interesting to consider the time evolution of these systems.  Firstly, since the quasi-steady surface density 
profile is independent of the initial conditions, continuing these simulations to $10^6$ years 
allows us to compare the quasi-steady nature for different disc masses. 
There is also no reason to suspect that the mass falling onto the disc will do so in such a way as to
immediately produce a quasi-steady disc.  We might therefore expect that the disc will not initially be in 
equilibrium. The settling times achieved here ($< 10^5$ years) are similar to and probably less than 
typical free-fall times \citep{wardthompson07}, suggesting that these systems could quite
quickly attain quasi-steady states.  Figure \ref{sdens} also shows that beyond $\sim 1$ au the disc reaches a 
quasi-steady state in 20000 years or less and might imply that these systems are rarely out of equilibrium. 

\subsection{Temperature profiles}

\begin{figure}
\begin{center}
\psfig{figure=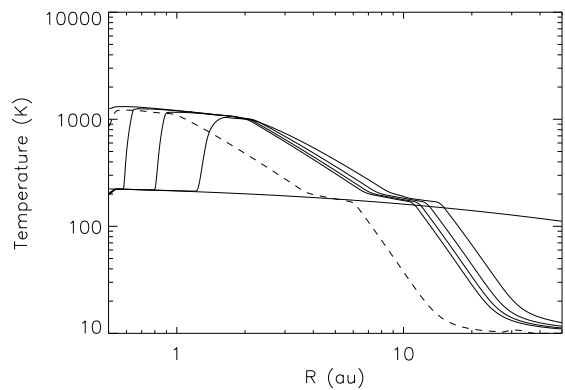,width=0.45\textwidth}
\caption{Disc Midplane temperature against radius for $\Sigma \propto r^{-1.5}$. The temperature
profiles for $\Sigma \propto r^{-1}$ and $\Sigma \propto r^{-2}$ aren't shown since the quasi-steady profiles
are the same.  The figure shows the initial temperature profile followed by 4 profiles separated
by 20000 years, and the profile after $10^6$ years (dashed line).  The form of the temperature
profile is essentially determined by the opacity.  Beyond $10$ au, ice grains dominate the opacity.
These melt at 170 K, producing the change in profile at $\sim 10$ au.  Within 10 au, metal grains dominate
the opacity until the temperature reaches $1000 - 2000$ K at which point these grains evaporate.}
\label{Tc}
\end{center}
\end{figure}

The evolution of the midplane temperature is shown in Figure \ref{Tc}.  Again, the Figure
shows the initial
temperature profile followed by $4$ subsequent temperature profiles separated by $20000$ years,
and a final temperature profile (dashed-line) after $10^6$ years.  Since the temperature is specified
using $Q = 1.5$ and $T_c \ge 10$ K, and since the quasi-steady profiles are essentially the same in all cases,
we show only the temperature profile for $\Sigma \propto r^{-1.5}$. 
From Figure \ref{Tc} and Figure \ref{kappa} we can start to understand the disc structure.  
The outer disc has $T_c < 170$ K and so the opacity is dominated by ice grains. The asymptotic 
nature of the temperature profile at large radii is simply due to the assumption that $T_c \ge 10$ K.
The change in slope
at, initially, a radius between $10$ and $20$ au that moves inwards with time is due to the 
evaporation of the ice mantles on the grains and is essentially the ``snowline". The opacity
of the inner part of the disc (within $\sim 10$ au) is dominated by metal grains until the temperature
reaches $1000 - 2000$ K where the grains evaporate and the temperature cannot rise significantly 
due to the sudden decrease in opacity. This results in a further change in the surface density
profile. The quasi-steady nature is therefore quite strongly dependent on the opacity of the
disc material. In the vertically averaged approximation
the optical depth is evaluated using only the midplane
value of the opacity. In the hot inner regions where the
midplane temperature exceeds the dust sublimation
temperature this is likely to result in an underestimation
of the true optical depth, since the surface regions are
cooler and could have a larger opacity. To evaluate the
true optical depth in detail would require solving for the
vertical structure and considering the time scales for the
evaporation and condensation of particulates within the turbulent
flow, which we do not attempt. Clearly, however, the sense
of the effect will be to increase the optical depth and
cooling time still further and hence the disc temperature
at the midplane will remain above the grain evaporation
temperature.

As will be discussed in more detail later, the $\alpha$ values in the inner regions of these
discs are low and, as shown in Figure \ref{alpha}, within $\sim 3$ au
are $< 10^{-4}$.  It is often assumed that 
$\alpha \sim 10^{-2}$ which, for mass accretion rates similar to those shown here, would imply much lower
temperatures in the inner disc.  Crystalline silicates - which are commonly observed in 
protostellar discs \citep{bouwman01,vanboekel04} and require $T > 800$ K - are therefore generally
thought to form close to the central star ($r \sim 0.1$ au for TTauri stars)
\citep{dullemond06} and then transported outwards via turbulent mixing
to at least 10 - 20 au \citep{gail01}.  Figure \ref{Tc}, however, suggests that
when the transport processes are dominated by self-gravity, the
temperature could be high enough to 
form crystalline silicates at radii of a few au.     

\subsection{Mass Accretion}
\begin{figure}
\begin{center}
\psfig{figure=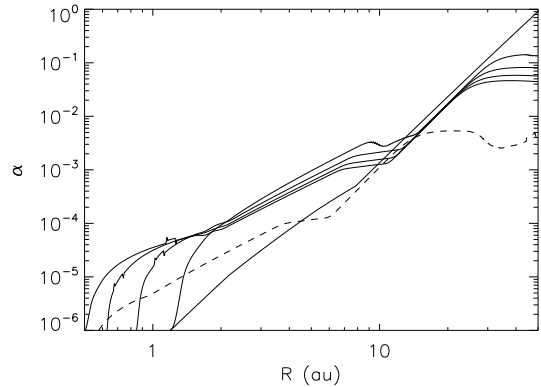,width=0.45\textwidth}
\caption{Viscous $\alpha$ against radius at different times for $\Sigma \propto r^{-1.5}$.
The results for $\Sigma \propto r^{-1}$ and $\Sigma \propto r^{-2}$ are essentially the
same, so aren't shown.  The figure shows the initial $\alpha$ profile followed by 4
$\alpha$ profiles separated by $20000$ years (solid lines) until a quasi-steady state, 
and the $\alpha$ profile after $10^6$ years (dashed line).  It has been
shown that fragmentation can occur if $\alpha \sim 0.06$. The $\alpha$ values in 
the inner disc can be very small, well below that required for fragmentation. The 
$\alpha$ value in the outer disc can, however, become quite large ($> 10^{-2}$) but in this
case is still below - but close to - that required for fragmentation.}
\label{alpha_fig}
\end{center}
\end{figure}

As discussed above (and illustrated in Figure \ref{Mdot_fig}), the disc relatively quickly
settles into a quasi-steady state with an approximately constant mass transfer rate.  The 
corresponding $\alpha$ values are shown in Figure \ref{alpha_fig}.  Again this Figure shows
the initial $\alpha$ profile, 4 subsequent $\alpha$ profiles separated by 20000 years,
and the final $\alpha$ profile after $10^6$ years.  The strong
dependence of $\alpha$ on the midplane temperature means that small
variations in $T_c$ can produce large variations in $\alpha$.  This
only really occurs in the outer disc when $T_c$ is close to the
minimum temperature of $10$ K, or where opacity changes produce
sudden changes in $T_c$ coupled with the initial conditions being far from the
equilibrium conditions.  The $\alpha$ and $\dot{M}$ profiles
have therefore been averaged to remove these
small-scale variations.  What Figure \ref{alpha_fig} illustrates is that the $\alpha$
value in a quasi-steady, self-gravitating disc is not constant and can reach extremely low-values
in the inner disc. In a more complete model other sources of transport (such as weak 
turbulence in the disc mid-plane excited by MRI active zones at the disc surface) might 
well dominate in this region. Such a process could also result in episodic accretion events
if the other source of transport were to depend on the disc properties - such as temperature
in the case of MRI -
and could explain phenomena such as FU Orionis outbursts \citep{hartmann96}.  
The mass accretion rate {\em into} the inner disc and ultimately onto the star, however, would 
still be set by the self-gravitating transport taking place at larger radii \citep{boley08}.

Figure \ref{Mdot_fig} also shows that for 
our chosen parameters ($M_* = 0.35$ M$_\odot$ and $M_{\rm disc} = 0.35$ M$_\odot$) the initial
quasi-steady mass accretion rate is $\sim 2 \times 10^{-7}$ M$_\odot$/yr.  The dashed line in Figure 
\ref{Mdot_fig} shows the accretion rate after $10^6$ years. At this stage, 
the disc mass is $\sim 0.25$ M$_\odot$ and the accretion rate is $\sim 10^{-8}$ M$_\odot$/yr. 
Since this is the quasi-steady rate for a $0.25$ M$_\odot$ disc around a $0.35$ M$_\odot$ star,
this illustrate that the accretion rate depends non-linearly on the disc mass (a factor of
1.4 reduction in disc mass reduces the accretion rate by about an order of magnitude).

This dependence of the quasi-steady accretion rate on the disc mass can be roughly understood
by considering the method discussed in section 2.  For a constant Toomre stability parameter,
$Q$, and central mass, the sound speed, $c_s$, depends linearly on the surface density.
Since $c_s \propto \sqrt{T_c}$, this gives $T_c \propto \Sigma^2$.  Using $\nu = \alpha c_s^2 / \Omega$, 
the mass accretion rate - shown in equation (\ref{Mdot}) - can be rewritten as $\dot{M} \propto
\alpha \Sigma^3$.  Combining equations (\ref{inten}), (\ref{alpha}), (\ref{tau}), and
(\ref{modCoolFunc}), and using $\tau \approx \kappa \Sigma$, gives $\alpha \propto \Sigma^4$ and consequently $\dot{M} \propto 
\Sigma^7$.  Here, for simplicity,  we ignore $T_o$ in equation (\ref{modCoolFunc}) and assume that
the disc is optically thick so $\tau / (1 + \tau^2) \approx \tau$. 
As $\Sigma$ is essentially determined by the disc mass, the accretion 
rate then depends on the disc mass to the seventh power.  A factor of 1.4 reduction in disc
mass would produce an order of magnitude change in accretion rate, consistent with 
that shown in Figure \ref{Mdot_fig}.  

\subsection{Fragmentation}

It has been shown \citep{gammie01,rice03} that there is a minimum cooling time, $t_{\rm cool}$, for which
a self-gravitating disc can remain in a quasi-steady state without fragmenting to form bound
objects.  The exact value of the minimum cooling time depends on the equation of state 
\citep{rice05} with fragmentation occuring for $t_{\rm cool} \le 3 \Omega^{-1}$ when
the specific heat ratio $\gamma = 5/3$ \citep{gammie01}.  \citet{rice05}, however, show that for all values
of $\gamma$ the fragmentation boundary occurs for $\alpha \sim 0.06$. Figure \ref{alpha_fig}
shows that once the disc settles into a quasi-steady state, the $\alpha$ values in the inner
disc ($r < 10$ au) are well below that required for fragmentation. Between $20$ and $50$ au,
however, $\alpha$ is almost constant and has a value ($\alpha \sim 0.04$) just below that required
for fragmentation. A small increase in disc mass would, however, increase $\alpha$ and could
produce conditions that would allow for fragmentation in the outer parts of such discs 
\citep{stamatellos07a, greaves08}.  We should, however, add that the results obtained
by \citet{gammie01} and \citet{rice03} were for systems in which the cooling time was chosen to be the same 
at all radii.  \citet{mejia05} argue that in systems with a radially varying cooling time, the
fragmentation boundary may not be a simple function of cooling time and that there could be regions
with $\alpha > 0.06$ that do not fragment.

\subsection{Variation with central mass}

\begin{figure}
\begin{center}
\psfig{figure=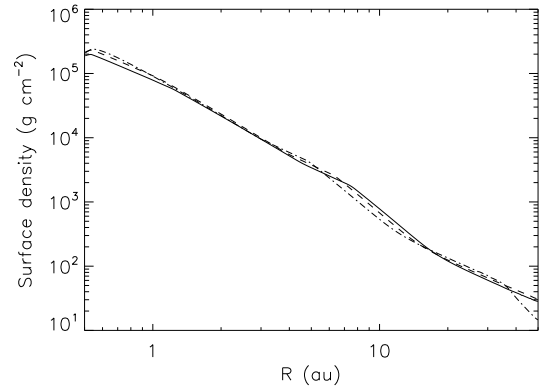, width=0.45\textwidth}
\caption{Quasi-steady surface density profiles for $M_{\rm disc} = 0.25$ M$_\odot$ and
$M_* = 0.25$ M$_\odot$ (solid line), $M_* = 0.35$ M$_\odot$ (dashed line), and $M_* = 0.5$
M$_\odot$ (dash-dot line).  The profiles are almost the same, but with a few differences
due to the different temperature profiles resulting from the Toomre $Q$ parameter's dependence on the
mass of the central star.}
\label{combsdens}
\end{center}
\end{figure}

\begin{figure}
\begin{center}
\psfig{figure=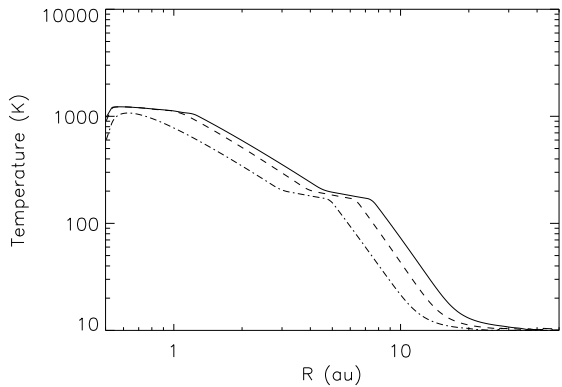, width=0.45\textwidth}
\caption{Quasi-steady temperature profiles for $M_{\rm disc} = 0.25$ M$_\odot$ and
$M_* = 0.25$ M$_\odot$ (solid line), $M_* = 0.35$ M$_\odot$ (dashed line), and $M_* = 0.5$
M$_\odot$ (dash-dot line).  The difference in the profiles is a result of the dependence of the 
Toomre $Q$ on the mass of the central star (through $\Omega$).  As the central star mass 
decreases, the sound speed - and hence temperature - needs to increase to maintain a constant 
$Q$ value.  This means that the radii at which ice mantles and metals evaporate differs
for the different central star masses, and produces (as seen in Figure \ref{combsdens})
slightly different surface density profiles.}
\label{combTc}
\end{center}
\end{figure}

We have also considered how the mass of the central object influences the quasi-steady nature of a 
self-gravitating disc.  To do this we consider a disc mass of $0.25$ M$_\odot$ around central
objects with masses of $M_* = 0.25$ M$_\odot$, $M_* = 0.35$ M$_\odot$, and $M_* = 0.5$ M$_\odot$.  In all
cases we assume an initial surface density profile of $\Sigma \propto r^{-2}$ and evolve the
system until a quasi-steady state is reached.  The resulting quasi-steady surface density profiles
are shown in Figure \ref{combsdens} with the solid line for $M_* = 0.25$ M$_\odot$, the
dashed line for $M_* = 0.35$ M$_\odot$, and the dashed-dot line for $M_* = 0.5$ M$_\odot$.  
The surface density profiles are, as expected, quite similar. The slight differences are a 
consequences of the dependence of the Toomre $Q$ parameter on the mass of the central object.  This can be
understood by considering Figure \ref{combTc} which shows the temperature profile for
the three cases.  Since we assume $Q = 1.5$, as the central object mass decreases 
(decreasing the orbital frequency $\Omega$) the sound speed, $c_s$, and hence
temperature, needs to increase. This changes the locations where the ice mantles
melt and where the metal grains evaporate and hence results in different central object masses
producing slightly different - although similar - surface density profiles.

\begin{figure}
\begin{center}
\psfig{figure=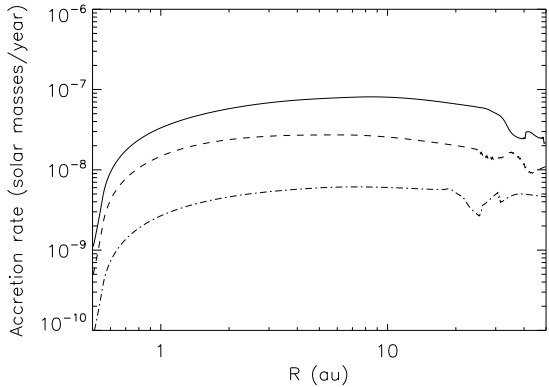, width=0.45\textwidth}
\caption{Quasi-steady mass accretion rates for $M_{\rm disc} = 0.25$ M$_\odot$ and
$M_* = 0.25$ M$_\odot$ (solid line), $M_* = 0.35$ M$_\odot$ (dashed line), and $M_* = 0.5$
M$_\odot$ (dash-dot line).  The accretion rate for $M_* = 0.25$ M$_\odot$ is significantly 
great than for $M_* = 0.5$ M$_\odot$. This is a consequence of the increased disc 
temperature required to keep the disc around the lower mass star gravitationally stable.
The strong dependence of $\alpha$ on temperature results in the $M_* = 0.25$ M$_\odot$ system
having an accretion rate more than an order of magnitude greater than the $M_* = 0.5$ M$_\odot$
system.}
\label{combMdot}
\end{center}
\end{figure}

Figure \ref{combMdot} shows the quasi-steady mass accretion rates for $M_* = 0.25$
M$_\odot$ (solid line), $M_* = 0.35$ M$_\odot$ (dashed line) and $M_* = 0.5$ M$_\odot$ (dash-dot line).
There is a very strong dependence on the central mass with the accretion rate for
$M_* = 0.25$ M$_\odot$ being more than an order of magnitude greater than that for
$M_* = 0.5$ M$_\odot$.  Again, we can understand this dependence by considering 
the method described in section 2.  For a constant disc mass, and hence surface 
density ($\Sigma$), a constant Toomre $Q$ requires $c_s \propto 1/\Omega \propto 1/\sqrt{M_*}$ giving
$T_c \propto 1/M_*$.  Using $\nu = \alpha c_s^2 / \Omega$, the mass accretion
rate can be rewritten as $\dot{M} \propto \alpha T_c^{3/2}$.   Equations (\ref{inten}), (\ref{alpha}), (\ref{tau}), and
(\ref{modCoolFunc}) can again be used to show that $\alpha \propto T_c^{5/2}$, giving
a mass accretion rate of $\dot{M} \propto T_c^{4}$, or $\dot{M} \propto 1/M_*^{4}$.
A factor of 2 increase in the mass of the central star therefore reduces the mass accretion
rate by a factor of 22, consistent with that seen in Figure \ref{combMdot}.  If, however,
we combine this relationship ($\dot{M} \propto 1/M_*^{4}$) with the relationship between
accretion rate and disc mass ($\dot{M} \propto \Sigma^7 \propto M_{\rm disc}^7$),
the accretion rate actually increase as $\dot{M} \propto M_*^3$ if a constant disc
to star mass ratio is maintained. This is consistent with the
observations 
that in the Orion Nebula cluster massive discs are more common around
lower-mass
stars \citep{eisner08}.

Something to point out is that as the disc mass decreases, it become increasingly difficult
to sustain steady, self-gravitating accretion in the outer regions of
the disc.  The surface density (Figures \ref{sdens} and
\ref{single_profile}) and 
temperature profiles (Figures \ref{combTc} and
\ref{Tc}) show that most of the mass is in the inner disc and that, as the disc mass decreases, the temperature in the outer
disc approaches $10$ K, our assumed minimum temperature.  Equation (\ref{modCoolFunc}) 
shows that as $T_c$ approaches $10$ K, the cooling function becomes very small, the 
cooling time becomes very long, and the effective $\alpha$ becomes very small. Decreasing
the disc mass much below $0.25$ M$_\odot$ effectively turns off self-gravitating accretion
beyond $\sim 30$ au.  This is, of course, due to our assumption that the quasi-steady
Toomre $Q$ value is 1.5. 
If quasi-steady transport can occur for larger values of $Q$,
self-gravitating transport could still operate effectively at lower
disc masses. This also has implications for the production of brown
dwarfs
and low-mass companions at large radii \citep{stamatellos07a},
suggesting that fragmentation at large radii is unlikely when the disc
is in a quasi-steady state. This doesn't proclude fragmentation at
large
radii, but implies that it must either occur when the disc is
extremely
massive, or must occur when the disc has not yet reached a
quasi-steady state. 

\begin{figure}
\begin{center}
\psfig{figure=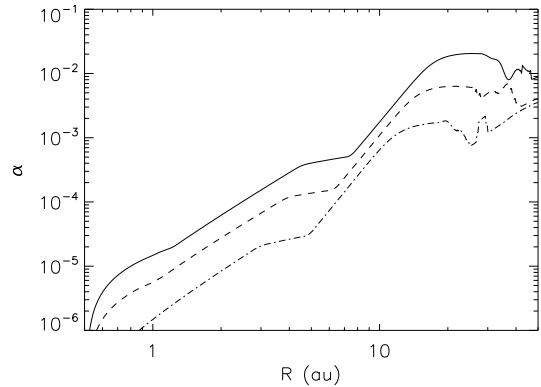, width=0.45\textwidth}
\caption{Effective viscous $\alpha$ profiles for $M_{\rm disc} = 0.25$ M$_\odot$ and
$M_* = 0.25$ M$_\odot$ (solid line), $M_* = 0.35$ M$_\odot$ (dashed line), and $M_* = 0.5$
M$_\odot$ (dash-dot line).  As discussed in the text, for a fixed disc
mass $\alpha$ increases with decreasing star mass.  In all cases,
however, the $\alpha$ values in the inner disc are well below that
required for fragmentation.  Although larger in the outer disc, the
values are close to, but still below, that needed for fragmentation.}
\label{combalpha}
\end{center}
\end{figure}

The corresponding $\alpha$ profiles are shown in Figure \ref{combalpha}.  The
dependence of $\alpha$ on $M_*$ discussed above gives larger $\alpha$ values
for $M_* = 0.25$ M$_\odot$ than for $M_* = 0.5$ M$_\odot$.  Again the $\alpha$
values in the inner disc are well below those required for fragmentation, while those
in outer disc are close to, but below, the required value ($\alpha = 0.06$).  As discussed earlier, however, small changes in some of the parameters
could produce conditions suitable for fragmentation in the outer parts of the disc, but
as suggested by \citet{rafikov05}, it is extremely difficult to see how fragmentation can occur
within $10$ au even for the relatively massive discs considered here.

\section{Discussion and Conclusions}
We consider here the evolution of protostellar discs in which angular
momentum transport is driven only by self-gravity.  We are particularly
interested in the quasi-steady nature of such systems, which we define here
as a state in which the mass accretion rate is approximately the same at all
radii. We assume that the disc will cool to a state of marginal 
gravitational stability ($Q \sim 1.5$) and will then be 
in thermal equilibrium.  The evolution of such systems is determined by the
effective gravitational viscosity which is calculated from the
assumption that the viscosity dissipates energy at the same rate as energy is lost from
the disc via radative cooling - which depends on the chosen
opacity.  This is in some sense the inverse
of a `` classical'' $\alpha$ disc \citep{shakura73} in which the
prescribed $\alpha$ value determines the viscosity, which then
sets the dissipation and
cooling rates.

We carried out a number of different simulations, initially
considering a central object of $0.35$ M$_\odot$ surrounded by an equal-mass disc with a
power-law surface density profile from $0.5$ au to
$50$ au.  We also considered central masses of $0.25$ M$_\odot$ and
$0.5$ M$_\odot$ with a disc mass of $0.25$ M$_\odot$. The discs are in
all cases evolved until a quasi-steady state is reached.  
The primary results are summarised below.

\begin{itemize}
\item{The discs reasonably quickly settle into a quasi-steady state
that is largely independent of the initial surface density profile.  The settling 
time does depend somewhat on the initial profile, but beyond $\sim 1$
au
a quasi-steady state is reached in $20000$ years or less, 
suggesting that such systems will rarely be out of equilibrium.}
\item{The quasi-steady
surface density profile is relatively steep ($\Sigma \sim r^{-2}$)
with changes in profile that correspond to the melting of 
ice grains (effectively the ``snowline'') and the evaporation of 
metal grains. The steep profile  
means that most ($\sim 80$ \%) of the mass is located 
within $10 - 20$ au, even though the discs extend initially to $50$
au.  Having a lot of the mass in the inner disc could aid planet formation, and is consistent with current 
observations showing that massive discs have small radii
\citep{rodriguez05}.}
\item{The corresponding temperature profile indicates that in a
  quasi-steady state the inner disc (within $\sim 3$ au) can be hot
  ($T_c > 1000$ K) which suggests that such regions could be sufficiently
  ionised to be unstable to the growth of MRI \citep{balbus91}.  The MRI viscosity
  is likely to be significantly larger than the effective gravitational
  viscosity, so will drain the inner disc, potentially producing FU
  Orionis-like outbursts \citep{armitage01,zhu09}. This process could
  repeat once the inner disc has been replenished by material from the
  outer disc. The hot inner disc ($T_c > 800$ K) also suggests that the formation of
  crystalline silicates could occur at relatively large radii ($r < 3
  - 4$ au) during
  the earliest stages of star formation.}
\item{The mass accretion rate in quasi-steady discs also has a strong dependence on disc mass 
  and on the mass of the central star.  From the simulations, and from
  analytic approximations, we find $\dot{M} \propto \Sigma^7 \propto
  M_{\rm disc}^7$ and $\dot{M} \propto 1/M_*^4$.  Together this
  suggests that for a constant disc to star mass ratio $\dot{M}
  \propto M_*^3$, consistent with suggestions \citep{kratter08} that
  the gravitational instability is the most important mechanism for
  stars with final masses $> 1$ M$_\odot$. This is also
  intriguingly similar to the $\dot{M} \propto M_*^2$ relationship
  between accretion rate and stellar mass determined observationally
  \citep{muzerolle05}, although the simulation results probably apply
  to systems considerably younger than those observed.}
\item{It has been shown \citep{gammie01, rice03} that
  fragmentation requires rapid cooling and is equivalent to $\alpha
  \le 0.06$ \citep{rice05}.  Despite the discs considered here being
  massive, none of the quasi-steady discs had $\alpha > 0.06$ at any radii.
  In the outer disc ($r > 20$ au) the $\alpha$ value was close to the
  fragmentation limit, but in the inner disc it was orders of
  magnitude below that required. Small changes in
  some disc properties could lead to fragmentation conditions at large
  radii \citep{stamatellos07a,greaves08}, but this does not appear to be
  possible in the inner disc, consistent with other calculations
  \citep{rafikov05,rafikov07,whitworth06}.  The quasi-steady surface density
  profile also has most of the mass in the inner disc,
  suggesting that if discs do fragment at large radii they either do
  so prior to settling into a quasi-steady state, or do so when the
  disc is extremely massive.}
\end{itemize}

Overall, the results presented here suggest a scenario in which -
while mass is falling from the envelope onto the disc - the
disc mass remains comparable to the central star mass and the system evolves towards a 
quasi-steady state with mass accretion rates in excess of $10^{-7}$
M$_\odot$/yr. MHD turbulence (MRI) will probably also play a role in some regions of the disc \citep{armitage01,terquem08} and,
in particular, could lead to FU Orionis outbursts if the inner disc temperature exceeds $\sim 1400$ K
\citep{armitage01,zhu09}.  Once envelope infall ceases, the disc mass
will start to decrease, rapidly reducing the mass accretion rate due to self-gravity.
The midplane in 
the inner disc will have a very low viscosity ($\alpha < 10^{-4}$) and
will effectively become dead.  This can occur for
relatively large disc masses ($M_{\rm disc} \sim 0.1$ M$_\odot$) which, together with most of the mass being in the inner 
disc, could enhance planet
formation.  The low viscosities in the midplane of the inner disc also means that embedded planets can easily
clear a large inner gap \citep{syer95} and could explain some of the observed systems with near-infrared deficits 
\citep{hartmann08}, often referred to as transition systems.  

\section*{acknowledgements}
P.J.A. acknowledges support from the NSF (AST-0807471), from
NASA's Origins of Solar Systems program (NNX09AB90G), and from NASA's
Astrophysics Theory and Fundamental Physics program (NNX07AH08G).
W.K.M.R. acknowledges support from the Scottish Universities Physics Alliance (SUPA).
The authors would also like to acknowledge useful discussions with Lee
Hartmann and Dick Durisen, and would like to thank the referee for their
constructive comments.


\begin{thebibliography}{}

\bibitem[Armitage, Livio \& Pringle(2001)]{armitage01}
 Armitage P.J., Livio M., Pringle J.E., 2001, MNRAS, 324, 705

\bibitem[Balbus \& Hawley(1991)]{balbus91}
 Balbus S.A., Hawley J.F., 1991, ApJ, 376, 214

\bibitem[Balbus \& Papaloizou(1999)]{balbus99}
 Balbus S.A., Papaloizou J.C.B., 1999, ApJ, 521, 650

\bibitem[Bell \& Lin(1994)]{bell94}
 Bell K.R., Lin D.N.C., 1994, ApJ, 427, 987

\bibitem[Blaes \& Balbus(1994)]{blaes94}
 Blaes O.M., Balbus S.A., 1994, ApJ, 421, 163

\bibitem[Boley \& Durisen(2008)]{boley08}
 Boley A.C., Durisen R.H., 2008, ApJ, 685, 1193

\bibitem[Boley et al.(2006)]{boley06}
 Boley A.C., Mej\'{\i}a A.C., Durisen R.H., Cai K., Pickett M.K., D'Alessio P., 2006, 651, 517

\bibitem[Boss(1998)]{boss98}
 Boss A.P., 1998, ApJ, 503, 923

\bibitem[Boss(2002)]{boss02}
 Boss A.P., 2002, ApJ, 576, 462

\bibitem[Boss(2004)]{boss04}
 Boss A.P., 2004, ApJ, 610, 456

\bibitem[Bouwman et al.(2001)]{bouwman01}
 Bouwman J., Meeus G., de Koter A., Hony S., Dominik C., Waters
 L.B.F.M., 2001, A\&A, 375, 950

\bibitem[Cai et al.(2006)]{cai06}
 Cai K., Durisen R.H., Michael S., Boley A.C., Mej\'{\i}a A.C., Pickett M.K. D'Alessio, P., 2006, ApJ, 636, L149

\bibitem[Carrasco-Gonzalez et al.(2009)]{carrasco-gonzalez09}
 Carrasco-Gonzalez C., Rodriguez L.F., Anglada G., Curiel S., 2009,
 ApJ, in press.

\bibitem[Caselli et al.(2002)]{caselli02}
 Caselli P., Benson P.J., Myers P.C., Tafalla M., 2002, ApJ, 572, 238

\bibitem[Eisner et al.(2008)]{eisner08}
 Eisner J.A., Plambeck R.L., Carpenter J.M., Corder S.A., Qi C.,
 Wilner D., 2008, ApJ, 683, 304

\bibitem[Dullemond, Apai \& Walch(2006)]{dullemond06}
 Dullemond C.P., Apai D., Walch S., 2006, ApJ, 640, L67

\bibitem[Durisen et al.(2007)]{durisen07}
 Durisen R.H., Boss A.P., Mayer L., Nelson A.F., Quinn T., Rice W.K.M., 2007,
 in Reipurth B., Jewitt D., Keil K., eds, Protostars and Planets V. University of Arizona
 Press, Tucson, p701

\bibitem[Forgan et al.(2009)]{forgan09}
 Forgan D., Rice K., Stamatellos D., Whitworth A., 2009, MNRAS, in press

\bibitem[Gail(2001)]{gail01}
 Gail H.-P., 2001, A\&A, 378, 192

\bibitem[Gammie(1996)]{gammie96}
 Gammie C.F., 1996, ApJ, 457, 355

\bibitem[Gammie(2001)]{gammie01}
 Gammie C.F., 2001, ApJ, 553, 174

\bibitem[Greaves et al.(2008)]{greaves08}
 Greaves J.S., Richards A.M.S., Rice W.K.M., Muxlow T.W.B., 2008, MNRAS, 391, L74

\bibitem[Haisch, Lada \& Lada(2001)]{haisch01}
 Haisch K.E., Lada E.A., Lada C.J., 2001, AJ, 121, 2065

\bibitem[Hartmann et al.(2006)]{hartmann06}
 Hartmann L., D'Alessio P., Calvet N., Muzerolle J., 2006, ApJ, 648, 484

\bibitem[Hartmann \& Kenyon(1996)]{hartmann96}
 Hartmann L., Kenyon S.J., 1996, ARA\&A, 34, 207

\bibitem[Hartmann(2008)]{hartmann08}
 Hartmann L., 2008, Physica Scripta, 130, pp. 014012

\bibitem[Hubeny(1990)]{hubeny90}
 Hubeny I., 1990, ApJ, 351, 632

\bibitem[Johnson \& Gammie(2003)]{johnson03}
 Johnson B.M., Gammie C.F., 2003, 597, 131

\bibitem[Kratter, Matzner \& Krumholz(2008)]{kratter08}
 Kratter K.M., Matzner C.D., Krumholz M.R., 2008, ApJ, 681, 375

\bibitem[Laughlin \& Bodenheimer(1994)]{laughlin94}
 Laughlin G., Bodenheimer P., 1994, ApJ, 436, 335

\bibitem[Laughlin \& Rocyzka(1996)]{laughlin96}
 Laughlin G., Rozycka M., 1996, ApJ, 456, 279

\bibitem[Lin \& Pringle(1987)]{lin87}
 Lin D.N.C., Pringle J.E., 1987, MNRAS, 225, 607

\bibitem[Lodato \& Rice(2004)]{lodato04}
 Lodato G., Rice W.K.M., 2004, MNRAS, 351, 630

\bibitem[Lodato \& Rice(2005)]{lodato05}
 Lodato G., Rice W.K.M., 2005, MNRAS, 358, 1489

\bibitem[Lynden-Bell \& Pringle(1974)]{lynden74}
 Lynden-Bell D., Pringle J.E., 1974, MNRAs, 168, 603

\bibitem[Matzner \& Levin(2005)]{matzner05}
 Matzner C.D., Levin Y., 2005, ApJ, 628, 817

\bibitem[Mejia et al.(2005)]{mejia05}
 Mej\'{\i}a A.C., Durisen R.H., Pickett M.K., Cai K., 2006, ApJ, 619, 1098

\bibitem[Muzerolle et al.(2005)]{muzerolle05}
 Muzerolle J., Luhmann K.L., Brice\~{n}o C., Hartmann L., Calvet N., 2005, ApJ, 625, 906

\bibitem[Najita et al.(2007)]{najita07}
 Najita J.R., Carr J.S., Glassgold A.E., Valenti J.A., 2007, in
 Reipurth B., Jewitt D., Keil K., eds, Protostars and Planets
 V. Univ. of Arizona Press, Tucson, p. 507 

\bibitem[Papaloizou \& Nelson(2003)]{papaloizou03}
 Papaloizou J.C.B., Nelson R.P., 2003, MNRAS, 339, 983

\bibitem[Pollack et al.(1996)]{pollack96}
 Pollack J.C.B., Hubickyj O., Bodenheimer P., Lissauer J.J., Podolak
 M., Greenzweig Y., 1996, Icarus, 124, 62

\bibitem[Pringle(1981)]{pringle81}
 Pringle J.E., 1981, ARA\&A, 19, 137

\bibitem[Rafikov(2005)]{rafikov05}
 Rafikov R.R., 2005, ApJ, 621, L69

\bibitem[Rafikov(2007)]{rafikov07}
 Rafikov R.R., 2006, ApJ, 648, 666

\bibitem[Rice et al.(2003)]{rice03}
 Rice W.K.M., Armitage P.J., Bate M.R., Bonnell I.A., 2003, MNRAS, 339, 1025

\bibitem[Rice et al.(2005)]{rice05}
 Rice W.K.M., Lodato G., Armitage P.J., 2005, MNRAS, 364, L56

\bibitem[Rodriguez et al.(2005)]{rodriguez05}
 Rodriguez L.F., Loinard L., D'Alessio P., Wilner D.J., Ho P.T.P.,
 2005, ApJ, 621, L133

\bibitem[Shakura \& Sunyaev(1973)]{shakura73}
 Shakura N.I., Sunyaev R.A., 1973, A\&A, 24, 337

\bibitem[Stamatellos et al.(2007a)]{stamatellos07a}
 Stamatellos D., Hubber, D.A., Whitworth A.P., 2007, MNRAS, 382, L30

\bibitem[Stamatellos et al.(2007b)]{stamatellos07b}
 Stamatellos D., Whitworth A.P., Bisbas T., Goodwin S., 2007, A\&A, 475, 37

\bibitem[Stamatellos \& Whitworth(2008)]{stamatellos08}
 Stamatellos D., Whitworth A.P., 2008, A\&A, 480, 879

\bibitem[Stamatellos \& Whitworth(2009)]{stamatellos09}
 Stamatellos D., Whitworth A.P., 2009, MNRAS, 392, 413

\bibitem[Syer \& Clarke(1995)]{syer95}
 Syer D., Clarke C., 1995, MNRAS, 277, 758

\bibitem[Terebey, Shu \& Cassen(1984)]{terebey84}
 Terebey S., Shu F.H., Cassen P., 1984, ApJ, 286, 529

\bibitem[Terquem(2008)]{terquem08}
 Terquem C.E.J.M.L.J., 2008, ApJ, 689, 532

\bibitem[Thi et al.(2005)]{thi05}
 Thi W.-F., van Dalen B., Bik A., Waters L.B.F.M., A\&A, 430, L61

\bibitem[Toomre(1964)]{toomre64}
 Toomre A., 1964, ApJ, 139, 1217.

\bibitem[van Boekel et al.(2004)]{vanboekel04}
 van Boekel R., et al., 2004, Nature, 432, 479 

\bibitem[Vorobyov \& Basu(2007)]{vorobyov07}
 Vorobyov E.I., Basu S., 2007, MNRAS, 381, 1009

\bibitem[Ward-Thompson et al.(2007)]{wardthompson07}
 Ward-Thompson D., Andr\'e P., Crutcher R., Johnstone D.,
Onishi T., Wilson C., 2007, in Reipurth B., Jewitt D., Keil L., eds,
Protostars and Planets V: An Observational Perspective of
Low-Mass Dense Cores II: Evolution Towards the Initial Mass Function.
Univ. of Arizona Press, Tucson, p.33

\bibitem[Whitworth \& Stamatellos(2006)]{whitworth06}
 Whitworth A.P., Stamatellos D., 2006, 458, 817

\bibitem[Zhu, Hartmann \& Gammie(2009)]{zhu09}
 Zhu Z., Hartmann L., Gammie C.F., 2009, ApJ, in press.

\end{thebibliography}
\end{document}